\documentclass[10pt,eps,twocolumn,superscriptaddress,showpacs,superscriptaddress]{revtex4-1}
\usepackage{url}
\usepackage{amsmath,amssymb}
\usepackage{fourier}

\usepackage[dvipdfm]{graphicx}
\usepackage{color}
\usepackage{setspace}

\usepackage{ulem}

\begin{document}
\title{Dynamical pattern formations in two dimensional fluid and Landau pole bifurcation}
\author{Shun Ogawa}
\email[E-mail: ]{sogawa@amp.i.kyoto-u.ac.jp}
\affiliation{
    Department of Applied Mathematics and Physics,
    Graduate School of Informatics, Kyoto University,
    606-8501 Kyoto, Japan}

\author{Julien Barr\'e}
\affiliation{
    Laboratoire J. A. Dieudonn\'e,
    Universit\'e de Nice Sophia-Antipolis,
    UMR CNRS 7351,
    Parc Valrose, F-06108 Nice Cedex 02, France
}

\author{Hidetoshi Morita}
\affiliation{Department of Mathematics, Kyoto University, 606-8502 Kyoto, Japan}
\affiliation{CREST, JST, 606-8502 Kyoto, Japan}

\author{Yoshiyuki Y. Yamaguchi}
\affiliation{
    Department of Applied Mathematics and Physics,
    Graduate School of Informatics, Kyoto University,
    606-8501 Kyoto, Japan}

\begin{abstract}
  A phenomenological theory is proposed to analyze the asymptotic
  dynamics of perturbed inviscid Kolmogorov shear flows in two
  dimensions. The phase diagram provided by the theory is in
  qualitative agreement with numerical observations, which include
  three phases depending on the aspect ratio of the domain and
  the size of the perturbation: a steady shear flow, a stationary
  dipole, and four traveling vortices.  The theory is based on a
  precise study of the inviscid damping of the linearized equation
and on an analysis of nonlinear effects.  In particular, we show that the dominant Landau
  pole controlling the inviscid damping undergoes a bifurcation, which
  has important consequences on the asymptotic fate of the
  perturbation.
\end{abstract}

\pacs{47.54.-r,47.15.ki,05.45.-a}
\maketitle

\section{Introduction}

Patterns in effective two-dimensional (2D) fluids flows are found in
nature in various contexts \cite{FB12, NJB13}: atmospheric
\cite{PSM88, BM11} and oceanic flows \cite{AV11} are examples.
To understand such large scale patterns
theoretically, the 2D Euler equation describing perfect fluids flows
is a simplified starting point. In this context, the study of
nonlinear structures, such as Kelvin's cat's eyes, over a shear flow
has a long history \cite{WT80,HL32,JTS67}, which includes more recent
mathematical developments \cite{HMR86,ZL11}.

Statistical physics has often been invoked to explain
the formation of large vortices since Onsager \cite{LO49, GLE06}.
The Miller-Robert-Sommeria (MRS) theory \cite{JM90,JM92, RR91, RRJS91},
which constructs the microcanonical measure for 2D Euler flows
by taking all the invariants into account,
is a particularly successful achievement.
A difficulty in applying this theory in practice is the fact that
the 2D Euler equation has infinitely many invariants.
More importantly,
the theory assumes the vorticity field on the large scales
to be stationary,
and therefore cannot describe non-stationary asymptotic behaviors.
For these cases, the statistical physics approach should be
supplemented by a dynamical understanding of pattern formations.

In this paper, we consider a shear base flow to which a
perturbation is added, and propose a phenomenological approach to
the pattern formations by analyzing the Landau pole of the
linearized equation with its bifurcation,
and by taking into account
nonlinear effects.

Detailed numerical simulations of this situation are presented
in Ref. \cite{HM11}: starting from a Kolmogorov flow on the doubly
periodic domain $\mathbb{T}^2 = [0, 2\pi) \times [0, 2 \pi \Gamma)$, a
perturbation of size $\epsilon$ is added; depending on the parameters
$(\epsilon, \Gamma)$, the perturbation may fully damp, evolve into a
stationary dipole, or create four long lived traveling vortices. The
regions on the $(\epsilon, \Gamma)$ plane where the dipole or traveling
vortices appear are numerically investigated in Ref. \cite{HM11}, but a
theoretical understanding is lacking. In this article, we aim at
providing such a theoretical explanation, by exploiting the analogy
between the 2D Euler equation for the vorticity field and the Vlasov equation
for plasmas. Indeed, a similar phenomenon has been investigated in
one-dimensional plasmas, described by the Vlasov equation. A naive
linear theory predicts that perturbations added to stable stationary
states damp exponentially \cite{LDL45}. It is well known, however, that,
if the perturbation is large enough, nonlinear effects
come into play, prevent complete damping, and may create traveling
clusters \cite{TMO65,GM97,MB00,CL98}; see also Ref. \cite{JB09} for the
simpler cases of ferromagnetic and anti-ferromagnetic Hamiltonian
mean-field model. To be more precise, according to Refs. \cite{TMO65,JB09}, 
such a phenomenon occurs when the two following criteria are satisfied:
\begin{enumerate}
\item The Landau damping time scale
is longer than a nonlinear timescale, called ``trapping timescale''.
\item If several clusters are formed, they should be so small
that they do not overlap and that a nonlinear superposition
  approximation \cite{MB95} may work.
\end{enumerate}
Although
these criteria are in view of nonlinear dynamics, they can be expressed
using the dominant Landau pole
computed from the linearized Euler equation. 
Our goal is to draw the phase diagram
in the $(\epsilon, \Gamma)$ plane
by using the above two criteria in the context of the Euler
equation, that is, by combining the phenomenological nonlinear estimate
and the linear Landau damping theory. 

This article is organized as follows.
In Sec. \ref{sec:conditions},
we qualitatively analyze criteria 1 and 2,
and show that they are related to the imaginary and the real parts of
the dominant Landau pole
of the initial Kolmogorov flow.
We therefore briefly review the linear theory for 2D
incompressible and inviscid fluids, and derive the dispersion relation
in Sec. \ref{sec:linear};
the computation of the
Landau pole requires an analytic continuation procedure similar to the
one used in Ref. \cite{RLS97}.
Using these computations,
we draw a phase
diagram in the $(\epsilon, \Gamma)$ plane in Sec. \ref{sec:phased}, and
compare this phenomenological estimate with numerical simulations in
Sec. \ref{sec:num}.  Section \ref{sec:sum} is devoted to the summary
and discussions.

\section{The criteria to observe a dipole, or traveling vortices}
\label{sec:conditions}

We start from the 2D Euler equation in the domain $\mathbb{T}^2$
\begin{equation}
    \label{eq:Eeq}
    \frac{\partial \omega}{\partial t} + \vec{v} \cdot \nabla \omega = 0,
\end{equation}
where the vorticity field $\omega$ and the velocity field $\vec{v}$
are related to a stream function $\psi$ through
\begin{equation}
    \label{eq:stream}
        \omega = \nabla^2 \psi, \quad
        \vec{v} = \left(-\frac{\partial \psi}{\partial y}, \frac{\partial\psi}{\partial x}\right).
\end{equation}
The periods for $x$ and $y$ axes are set as $2\pi$ and $2\pi\Gamma$ respectively.

We consider the stationary Kolmogorov flow, called hereafter the ``base
flow'',
whose stream function is $\psi_{0}=-\Gamma\sin(y/\Gamma)$ and
vorticity and velocity fields are respectively
\begin{equation}
    \label{eq:base}
    \omega_{0}(y) = -U'(y), \quad
    \vec{v}_{0} = (U(y),0), \quad
\end{equation}
with $U(y)=\cos(y/\Gamma)$.
We add a large scale 
perturbation to the base flow, and expand the stream function as
\begin{equation}
    \psi(x,y,t) = \psi_{0}(y) + \psi_{1}(x,y,t),
\end{equation}
with
\begin{equation}
\label{eq:perturbation}
\psi_1(x,y,t=0) = \epsilon \cos x.
\end{equation}
In our analytical computations, $\epsilon$ is assumed to be small;
this restriction obviously does not hold for numerical simulations.
Nevertheless, the theory will be qualitatively in good agreement with
the simulations.

When the base flow is stable (this corresponds to an aspect ratio
$\Gamma>1$), the linear theory typically predicts that the
perturbation damps and possibly oscillates at complex frequency
$c=c_{\rm R}+ic_{\rm I}$,
where $c$ is the root of the dispersion relation yielding the slowest
damping, that is, the root closest to the real axis, or, in other
words, with the smallest $|c_{\rm I}|$. The idea is that while it damps, the
perturbation will tend to create structures traveling at velocity
$c_{\rm R}$; if nonlinear effects are strong enough, these structures
may persist for long times. If $c_{\rm R}\neq 0$, one may
expect
traveling vortices, while the formation of two stationary vortices
(a dipole) should be favored if $c_{\rm R}=0$.

\subsection{Criterion 1: damping timescale longer than trapping timescale}
The damping time scale $\tau_{\rm L}$ is easily estimated as the
inverse of the Landau damping rate $\tau_{\rm L} \simeq 1/|c_{\rm
  I}|$.  Actually, in addition to the exponential Landau damping
described by $c_{\rm I}$, there is an algebraic damping coming from the
branch points of the dispersion function (see Sec.~\ref{sec:linear});
however, the exponential Landau damping is expected to be dominant 
on the time scale considered in this paper.

Next, we estimate the ``trapping time scale'', $\tau_{\rm T}$,  which 
is the characteristic time scale concerning nonlinearities~\cite{TMO65},
as the period of a test point vortex trapped around the edge of a small 
vortex. The
temporal evolution of the position of a test point vortex is governed
by the velocity field
\begin{equation}
    \label{eq:EOM}
    \dot{x}= - \frac{\partial \psi}{\partial y}
    = U(y) - \frac{\partial \psi_{1}}{\partial y},
    \qquad
    \dot{y} = \frac{\partial \psi}{\partial x}
    =\frac{\partial \psi_{1}}{\partial x},
\end{equation}
where the dot denotes ${\rm d}/{\rm d}t$.
We approximate the perturbation $\psi_{1}(x,y,t)$
phenomenologically.  We are interested in the macroscopic behavior
corresponding to the $k=\pm 1$ mode, where $k$ is the wave number with respect 
to $x$,
and we assume that the damping
perturbation has created small vortices whose velocity is $c_{\rm R}$
in the $x$ direction; note that $c_{\rm R}$ can be ~$0$. The first
order $\psi_{1}(x,y,t)$ can be, therefore, approximated by
\begin{equation}
    \label{eq:psi1-approx}
    \psi_{1}(x,y,t) \simeq \epsilon \hat{\psi}_{1}(y)\cos (x-c_{\rm R}t),
    \quad U(y)\sim c_{\rm R},
\end{equation}
where $\hat{\psi}_{1}(y)$ is of $O(\epsilon^{0})$.
Substituting Eq. \eqref{eq:psi1-approx} into Eq. \eqref{eq:EOM},
we have the approximate equations of motion
\begin{equation}
    \label{eq:EOM01}
    \begin{split}
        \dot{x} &= U(y) - \epsilon \hat{\psi}'_{1}(y)\cos(x-c_{\rm R}t), \\
        \dot{y} &= - \epsilon \hat{\psi}_{1}(y) \sin(x -c_{\rm R} t),
    \end{split}
\end{equation}
where the prime represents ${\rm d}/{\rm d}y$.
Thanks to the phenomenological approximation of $\psi_{1}$
\eqref{eq:psi1-approx}, we can compute the motion of the test point
vortex by perturbation techniques.  Expanding $x$ and $y$ into series
of $\epsilon$ as
\begin{equation}
    \label{eq:xy}
    \begin{split}
        x(t) &= x_0(t) + \epsilon x_1(t) + O(\epsilon^2),
        \\
        y(t) &= y_0(t) + \epsilon y_1(t) + O(\epsilon^2),
    \end{split}
\end{equation}
we obtain the solutions $x_{0}(t)$ and $y_{0}(t)$ with initial
conditions $x_{0}(0)=X$ and $y_{0}(0)=Y$:
\begin{equation}
    \label{eq:x0y0}
    x_{0}(t) = X + U(Y)t, \quad
    y_{0}(t) = Y.
\end{equation}
Equations for the first order in $\epsilon$ are:
\begin{equation}
    \label{eq:EOM1}
    \begin{split}
        \dot{x}_1(t) &=
        U'(Y) y_1(t) - \hat{\psi}_{1}'(Y) \cos\left(X + (U(Y)-c_{\rm R})t
\right), \\
        \dot{y}_1(t) &=
        - \hat{\psi}_{1}(Y) \sin\left(X + (U(Y)-c_{\rm R})t \right).
    \end{split}
\end{equation}
From the frequency of $y_{1}$ in Eq. \eqref{eq:EOM1},
we estimate the trapping time scale $\tau_{\rm T}$ as
\begin{equation}
    \label{eq:tauT}
    \tau_{\rm T} \simeq \frac{1}{|U(Y_\ast + g(\epsilon, \Gamma)) - c_{\rm
R}|}
    \simeq \frac{1}{g(\epsilon, \Gamma) |U'(Y_\ast)|}
\end{equation}
where $Y_{\ast}$ satisfies $U(Y_{\ast}) = c_{\rm R}$,
and $g(\epsilon,\Gamma)>0$ represents the width of a vortex.

The unknown quantity, the width of a vortex $g(\epsilon,\Gamma)$,
is determined self-consistently.
The solution $y_{1}(t)$ is, from Eq. \eqref{eq:EOM1},
\begin{equation}
    y_{1}(t)
    = \dfrac{\hat{\psi}_{1}(Y)}{U(Y)-c_{\rm R}}
    \left[ \cos (X+(U(Y)-c_{\rm R})t) - \cos X \right],
\end{equation}
where the initial condition is $y_{1}(0)=0$.
The amplitude in the $y$ direction, 
$\epsilon\hat{\psi}_{1}(Y)/(U(Y)-c_{\rm R})$,
must be the same as the width $g(\epsilon,\Gamma)$
at the edge of the vortex, $Y=Y_{\ast}+g$.
Thus, the self-consistent equation for $g$ is
\begin{equation}
    g = \dfrac{\epsilon}{U(Y_{\ast}+g)-c_{\rm R}},
\end{equation}
where we have introduced another phenomenological approximation
by replacing $\hat{\psi}_{1}(Y)$ with $1$,
since it is of $O(\epsilon^{0})$ and we
are looking for an order-of-magnitude estimate. The width $g$ is,
therefore, estimated as
\begin{equation}
    \label{eq:g}
    g(\epsilon,\Gamma) \approx \sqrt{\dfrac{\epsilon}{|U'(Y_{\ast})|}}
\end{equation}
For the Kolmogorov base flow, $U(y)=\cos(y/\Gamma)$, we have
\begin{equation}
    \label{eq:UdashYast}
    |U'(Y_{\ast})| = \dfrac{1}{\Gamma} \sqrt{1-c_{\rm R}^2},
\end{equation}
and the trapping time scale is
\begin{equation}
    \label{eq:tauT-Kolmogorov}
    \tau_{\rm T} = \sqrt{\dfrac{\Gamma}{\epsilon}} 
    ( 1-c_{\rm R}^{2})^{-1/4}.
\end{equation}
Criterion 1 reads $\tau_{\rm T}<\tau_{\rm L}$, that is:
\begin{equation}
    \label{eq:cond1}
    \dfrac{\Gamma c_{\rm I}^{2}}{\sqrt{1-c_{\rm R}^{2}}}
    < \epsilon.
\end{equation}
If $c_{\rm R}=0$, this condition simplifies into 
$\Gamma c_{\rm I}^2<\epsilon$.

\subsection{Criterion 2: non overlapping vortices}
According to Eq. \eqref{eq:perturbation}, we consider the modes $k=\pm 1$.
We assume that the frequency $c_{\rm R}$ of the perturbation reflects
the existence of periodically moving vortices,
thanks to the periodic boundary condition for the $x$ axis.
The period $2\pi$ permits to identify the frequency
with the velocity of the moving vortices, and therefore
the $y$ positions of the vortices are estimated
as the solutions of the equation $U(y)=\pm c_{\rm R}$.
We recall the base flow is $U(y)=\cos(y/\Gamma)$.

If $c_{\rm R}=0$, one expects that vortices
are formed at $y=\pi \Gamma /2$ and $y=3 \pi\Gamma /2$.
The estimate of the vortices width \eqref{eq:g} shows that the vortices
  will never overlap for any reasonably small $\epsilon$ (say for
  instance $\epsilon<0.5$).
Hence criterion 2 brings no restriction in this case.

\begin{figure}
    \centering
    \includegraphics[width=7cm]{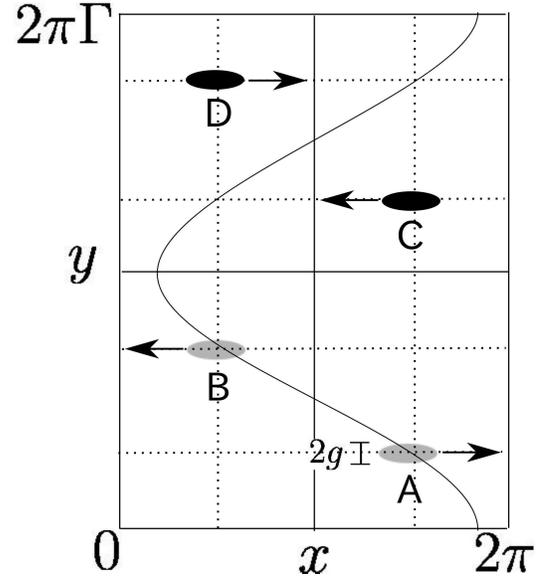}
    \caption{Schematic picture of four traveling vortices, A, B, C,
      and D. The arrows indicate traveling directions.  Vortices A and B
      have positive vorticity, while C and D negative.
The solid curve represents the velocity of the base flow $U(y)$, suitably 
rescaled.
The two vertical dotted lines 
      correspond to
      $\pm c_{\rm R}$; their intersections with the velocity
      curve, $U(y)=\pm c_{\rm R}$, yield the estimated $y$ positions
      of the four vortices, which travel along the horizontal dotted lines.
      }
    \label{fig:schematic}
\end{figure}

If $c_{\rm R}\neq 0$, in contrast,
the vortices can be close one to another,
and criterion 2 leads to a restriction.
We name the four vortices A, B, C, and D,
 whose $y$ positions are, respectively,
$\Gamma\arccos(|c_{\rm R}|)$, $\pi\Gamma-\Gamma\arccos(|c_{\rm R}|)$,
$\pi\Gamma+\Gamma\arccos(|c_{\rm R}|)$ and
$2\pi\Gamma-\Gamma\arccos(|c_{\rm R}|)$ in the interval $[0,2\pi\Gamma)$,
where we take the branch of solutions 
$0\leq \arccos(|c_{\rm R}|) \leq\pi/2$; see Fig.~\ref{fig:schematic}.

We assume that each vortex has the same width $g$.
Then the distance
between two nearby vortices must be larger than $2g$ in order to avoid
an overlap. Using $\pi/2-\arccos(c_{\rm R})=\arcsin(c_{\rm R})$, this
condition is expressed as
\begin{equation}
    \label{eq:2nd-necessary-condition-simple}
    \left\{
    \begin{array}{ll}
        \Gamma\arcsin(c_{\rm R}) > g, &
        ( |c_{\rm R}|<1/\sqrt{2} ) , \\
        \Gamma\arccos(c_{\rm R}) > g, &
        ( |c_{\rm R}|>1/\sqrt{2} ) . \\
    \end{array}
    \right.
\end{equation}
The first inequality comes from the distance in $y$
between vortices A and B (identically C and D), and the 
second from the distance between B and C (identically A and D).
The pair (B,C) moves toward the left while (A,D) toward the right
according to the base flow $U(y)$.
Moreover, the numerical observations \cite{HM11} say that
the difference in $x$ among each pair stays around $\pi$.
Thus the distance between B and C (A and D) is large enough 
to neglect the effect of overlapping; see Fig. \ref{fig:schematic}.
Hence we can omit the second inequality of Eq. (19),
and criterion 2 is finally expressed as:
\begin{equation}
    \label{eq:cond2}
    \epsilon < \Gamma \sqrt{1-c_{\rm R}^{2}} (\arcsin(c_{\rm R}))^{2}.
\end{equation}

The two conditions \eqref{eq:cond1} and \eqref{eq:cond2} involve the
parameters $\Gamma$ and $\epsilon$, as well as the dominant Landau
pole $c=c_{\rm R}+ic_{\rm I}$. In the following section, we turn to the
computation of this Landau pole.

\section{The linear theory and the dispersion relation}
\label{sec:linear}

This section contains a classical computation
for the linearized 2D Euler equation,
as well as an analytic continuation in the spirit of Ref. \cite{RLS97}, 
for the self-consistency of the paper.
We perform  it for a general base flow and perturbation,
and then specialize it to Eqs. \eqref{eq:base} and \eqref{eq:perturbation}.

\subsection{The dispersion function $D(c)$}

We linearize Eqs. \eqref{eq:Eeq} and \eqref{eq:stream} around the base
flow \eqref{eq:base}.
We add a small perturbation $\psi_{1}$ to $\psi_{0}(y)$, whose 
associated vorticity and velocity fields are
denoted respectively as $\omega_{1}$ and $\vec{v}_{1}$.
Substituting $\omega=\omega_{0}+\omega_{1}$ and $\vec{v}=\vec{v}_{0}+\vec{v}_{1}$
into the Euler equation \eqref{eq:Eeq}
gives the linearized 2D Euler equation
\begin{equation}
    \label{eq:lEeq}
    \frac{\partial \omega_1}{\partial t}
        + \vec{v}_0 \cdot \nabla \omega_1
        + \vec{v}_1 \cdot \nabla \omega_0 = 0.
\end{equation}
Using the perturbation $\psi_1$ and Eq. \eqref{eq:base}, we rewrite
the linearized Euler equation \eqref{eq:lEeq} as
\begin{equation}
    \label{eq:lEeqs}
    \frac{\partial \omega_1}{\partial t}
        + U(y) \frac{\partial \omega_1}{\partial x}
        - U''(y)\frac{\partial \psi_1}{\partial x} = 0.
\end{equation}
Our goal is now to derive the dispersion relation for the linearized
Euler equation; we use a Fourier-Laplace transformation, and follow
the route of Ref. \cite{FB10}.
Thanks to the periodicity with
respect to $x$, we expand $\omega_1$ and $\psi_1$ into Fourier series as
\begin{equation}
    \label{eq:F1}
    \begin{split}
        \omega_1(x,y,t) &=
        \sum_{k\in\mathbb{Z}} \hat{\omega}_k (y,t) e^{ikx},  \\
        \psi_1(x,y,t) &=
        \sum_{k\in\mathbb{Z}} \hat{\psi}_k (y,t) e^{ikx}.
    \end{split}
\end{equation}
Substituting the Fourier expansions \eqref{eq:F1} into
the linearized Euler equation \eqref{eq:lEeqs},
we obtain the equation for $k$-th Fourier mode as
\begin{equation}
    \label{eq:lEeqk}
    \frac{\partial \hat{\omega}_k}{\partial t} + ikU(y) \hat{\omega}_k
    - ik \hat{\psi}_k U''(y) = 0, \quad \forall k\in \mathbb{Z}.
\end{equation}
The Laplace transform of a function $\hat{g}(t)$ with respect to $t$
is defined as
\begin{equation}
    \tilde{g}(z) = \int_0^\infty \hat{g}(t) e^{-zt} dt,
    \quad {\rm Re}~z > 0,
\end{equation}
where the condition ${\rm Re}~z>0$ is introduced to ensure
the convergence of the integral.
Performing the Laplace transform of
the linearized Euler equation \eqref{eq:lEeqk},
we obtain an algebraic equation for $\tilde{\omega}_{k}$
and $\tilde{\psi}_{k}$:
\begin{equation}
    \label{eq:lEeqc}
    \left(U(y) - c\right) \tilde{\omega}_k - U''(y) \tilde{\psi}_k =
    \frac{\hat{\omega}_k(y, 0)}{ik},
\end{equation}
where $z=-ikc$; note that $\mathrm{Im}(kc)>0$.

The Fourier transform of Eq.~\eqref{eq:stream} with respect to $x$
\begin{equation}
    \label{eq:F2}
    \tilde{\omega}_{k}(y,c)
    = \left(\frac{\partial^2}{\partial y^2} - k^2\right)
\tilde{\psi}_{k}(y,c)
\end{equation}
gives a closed equation for $\tilde{\psi}_{k}(y,c)$,
which is the Rayleigh equation
\begin{equation}
    \label{eq:Req}
    \frac{\partial^{2} \tilde{\psi}_{k}}{\partial y^{2}}
        - k^{2} \tilde{\psi}_{k} -
    \frac{U''(y)}{U(y) -c} \tilde{\psi}_{k} =
    \frac{\hat{\omega}_{k}(y,0)}{ik (U(y) - c)}.
\end{equation}
The stream function $\hat{\psi}_k(y,t)$ is computed through an
inverse
Laplace transform. Its asymptotic behavior is determined
by the
singularities of $\tilde{\psi}_{k}(y,c)$
with respect to the complex variable $c$.
For simplicity, we introduce the functions $q$ and $f$ as
\begin{equation}
    \label{eq:function-q}
    q(y) \equiv k^2 +\frac{U''(y)}{U(y) -c}, \quad
    f(y) \equiv  \frac{\hat{\omega}_k(y,0)}{ik (U(y) - c)}.
\end{equation}
These functions have no singularity for real $y$,
since $c$ is defined in the region ${\rm Im}~(kc)>0$.

Fixing the complex variable $c$, we analyze
the Rayleigh equation of the form
\begin{equation}
    \label{eq:Req1}
    \frac{{\rm d}^2 \phi}{{\rm d}y^2} -q(y) \phi = f(y),
\end{equation}
and the corresponding homogeneous Rayleigh equation
\begin{equation}
    \label{eq:Reqh}
    \frac{{\rm d}^2 \phi}{{\rm d}y^2} -q(y) \phi = 0.
\end{equation}

Let $\phi_1$ and $\phi_2$ be independent solutions
to the homogeneous equation \eqref{eq:Reqh}
with boundary conditions
\begin{equation}
    \label{eq:bc1}
    \begin{cases}
        \phi_1(0) &= 1,\\ \phi_1'(0) &= 0,
    \end{cases}
    \quad \textrm{and}\quad
    \begin{cases}
        \phi_2(0)& = 0,\\ \phi_2'(0) &= 1.
    \end{cases}
\end{equation}
The particular solution $\phi_{\rm p}$ to the inhomogeneous equation
\eqref{eq:Req1} is then given by
\begin{equation}
    \begin{split}
    \phi_{\rm p}(y) = - \phi_1(y) & \int_{0}^{y} \phi_2(y') f(y')~dy' \\
    &+ \phi_2(y)\int_{0}^{y} \phi_1(y') f(y')~dy'.
    \end{split}
\end{equation}
Indeed, the double derivative of $\phi_{\rm p}(y)$ is
\begin{equation}
    \phi_{\rm p}''(y) = q(y) \phi_{\rm p}(y) + W(y) f(y),
\end{equation}
where $W(y)$ is the Wronskian
\begin{equation}
    W(y) = \phi_1(y)\phi_2'(y)-\phi_1'(y)\phi_2(y)
\end{equation}
and is constant
\begin{equation}
    \label{eq:Wronskian-constant}
    W(y)=W(0)=1 \quad \text{for all } y.
\end{equation}
Hence, the general solution $\phi_{\rm g}$ to Eq. \eqref{eq:Req} is:
\begin{equation}
    \label{eq:gsol}
    \phi_{\rm g} = \phi_{\rm p} + a_1 \phi_1 + a_2 \phi_2,
\end{equation}
where $a_1$ and $a_2$ are constants
determined from the
periodic boundary condition
\begin{equation}
    \label{eq:PBC}
    \phi_{\rm g}(2\pi\Gamma) = \phi_{\rm g}(0), \quad
    \phi_{\rm g}'(2\pi\Gamma) = \phi_{\rm g}'(0).
\end{equation}
From the boundary conditions \eqref{eq:bc1} and
$\phi_{\rm p}(0)=\phi_{\rm p}'(0)=0$,
Eqs.~\eqref{eq:gsol} and \eqref{eq:PBC} lead to:
\begin{equation}
    \label{eq:D}
    \begin{pmatrix}
    \phi_1(2\pi\Gamma)-1  & \phi_2(2\pi\Gamma)
    \\
    \phi_1'(2\pi\Gamma) & \phi_2'(2\pi\Gamma)-1
    \end{pmatrix}
    \begin{pmatrix}
    a_1 \\ a_2
    \end{pmatrix}
    = -
    \begin{pmatrix}
    \phi_{\rm p}(2\pi\Gamma) \\
        \phi_{\rm p}' (2\pi\Gamma)
    \end{pmatrix}.
\end{equation}
%\begin{equation}
%    \label{eq:D}
%    \begin{pmatrix}
%    \phi_1(\alpha+2\pi\Gamma)-1  & \phi_2(\alpha+2\pi\Gamma)
%    \\
%    \phi_1'(\alpha+2\pi\Gamma) & \phi_2'(\alpha+2\pi\Gamma)-1
%    \end{pmatrix}
%    \begin{pmatrix}
%    a_1 \\ a_2
%    \end{pmatrix}
%    = -
%    \begin{pmatrix}
%    \phi_{\rm p}(\alpha+2\pi\Gamma) \\
%        \phi_{\rm p}' (\alpha+2\pi\Gamma)
%    \end{pmatrix}.
%\end{equation}
Remembering that $q(y)$ and hence $\phi_{1}$ and $\phi_{2}$ depend
on $c$,
we define the function $D(c)$ as
\begin{equation}
    \label{eq:Dc}
    \begin{split}
        D(c) &= \det
        \begin{pmatrix}
            \phi_1(2\pi\Gamma)-1 &\phi_2(2\pi\Gamma)
            \\
            \phi_1'(2\pi\Gamma)   &\phi_2'(2\pi\Gamma) - 1
        \end{pmatrix}\\
        &=
        2- \phi_1(2\pi\Gamma)-\phi_2'(2\pi\Gamma).
    \end{split}
\end{equation}
To show the last equality in Eq. \eqref{eq:Dc}, we have used the fact
that the Wronskian $W(y)$ is unity.  The values of
$\phi_{1}(2\pi\Gamma)$ and $\phi_{2}'(2\pi\Gamma)$ are
computed by integrating the homogeneous Rayleigh equation
\eqref{eq:Reqh} from $0$ to $2\pi\Gamma$.

\subsection{Analytic continuation of $D(c)$}
\label{sec:extension}

The general solution \eqref{eq:gsol} has singularities for $c$
satisfying $D(c)=0$, and these singularities yield non-trivial modes
proportional to $e^{-ikct}$ by the inverse Laplace transform.
This justifies our terminology ``dispersion relation'' for $D(c)=0$, and
``dispersion function'' for the function $D(c)$.
Recall that the dispersion function $D(c)$ is a priori defined
only in the region ${\rm Im}(kc) > 0$
to ensure the convergence of the Laplace transform.
Now, to find roots giving stable modes,
we analytically continue $D(c)$ to the whole complex $c$ plane,
in a similar manner to what has been performed for the Euler
equation on the 2D disc \cite{RJB70, NRC95, RLS97}.

\begin{figure}
    \centering
    \includegraphics[width=7cm]{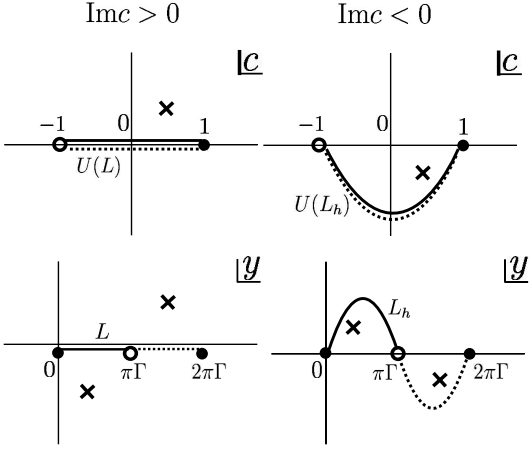}
      \caption{Schematic picture of complex $c$ and $y$ planes
        for $k=1$.
        The left two panels are for ${\rm Im}~c>0$,
        and the right two panels for ${\rm Im}~c<0$.
        The integration path $L$ on the $y$ plane,
        consisting of the solid and the dotted lines,
        is mapped to the double covering of the interval $[-1,1]$
        on the $c$ plane by the mapping $U:y\mapsto U(y)$.
        Corresponding to a given cross point on the $c$ plane,
          there are two singular points (crosses) of $q(y)$ and $f(y)$
          on the $y$ plane in general.
        As the point on the $c$ plane goes down,
      the singular points
      on the $y$ plane pass the real axis
        and the integration path $L$ must be deformed to $L_{h}$
        to avoid the singularities.
        The mapping of $L_{h}$, denoted by $U(L_{h})$,
        determines the boundary of
        the continued domain of the dispersion function $D(c)$.
        See Fig. \ref{fig:sp-deform}.
        A branch cut on the $c$ plane is set as the boundary.}
    \label{fig:continuation}
\end{figure}

Hereafter we set $k=1$ without loss of generality, and
take the particular form of the base flow: $U(y)=\cos(y/\Gamma)$.
Corresponding to a given $c$,
the functions $q(y)$ and $f(y)$ have singularities
at which the equation $U(y)-c=0$ is satisfied.
Such singular points are,
together with the integration paths of the homogeneous Rayleigh
equation \eqref{eq:Reqh},
schematically illustrated in Fig. \ref{fig:continuation}
for both the cases $\mathrm{Im}~c>0$ and $\mathrm{Im}~c< 0$.
The integration paths are determined as follows.

For $c$ on the upper half plane, $D(c)$ is simply defined by 
integrating Eq. \eqref{eq:Reqh} along the real $y$ axis, namely along the 
integration path:
\begin{equation}
    L = \left\{ y \in \mathbb{R} ~|~
      y = y(s) = 2\pi\Gamma s, ~ s\in [0,1) \right\}.
\end{equation}
Continuously moving $c$ to the lower half plane allows to define
$D(c)$ for $\mathrm{Im}~c<0$.
If the path taking $c$ from the upper to the lower 
half plane crosses the singular interval $c\in[-1,1]$, we have to avoid the 
singularity that would be created. For this purpose, we deform the 
integration path $L$ to
\begin{equation}
    \label{eq:ysh}
    L_{h}
    = \left\{ y \in \mathbb{C} ~|~
      y = y_{h}(s) = 2\pi \Gamma(s + i h(s)),~ s\in[0,1)
\right\}~,
\end{equation}
using a $C^2$-class, real-valued, and 1-periodic function $h(s)$
satisfying $h(0) = h(1) = 0$; correspondingly, this amounts to deform 
the singular line
\begin{equation}
    \sigma  = \{ c \in\mathbb{R} ~|~ c=U(y(s)),~ s\in [0,1) \},
\end{equation}
which doubly covers the interval $[-1,1]$,
to the curve
\begin{equation}
    \sigma_{h} = \{ c \in\mathbb{C} ~|~ c=U(y_{h}(s)),~ s\in [0,1) \}.
\end{equation}
We choose the function $h(s)$ so that the deformed curve $\sigma_{h}$
is below the $c$ for which we want to define $D(c)$.
This procedure is similar to the spectral deformation
for the Vlasov-Poisson equation in Refs. \cite{JDC89,PDH89}.

If the path taking $c$ from the upper to the lower plane crosses the
real axis through either one of the half lines $(-\infty,-1)$ or
$(1,\infty)$, no singularity is crossed, and the analytic continuation
does not require any deformation of the integral path; in the lower
half plane, $D(c)$ is multi valued
unless a Riemann surface associated with the branch
points $c=\pm 1$ is introduced.

\begin{figure}[t]
    \begin{center}
            \includegraphics[width=7cm]{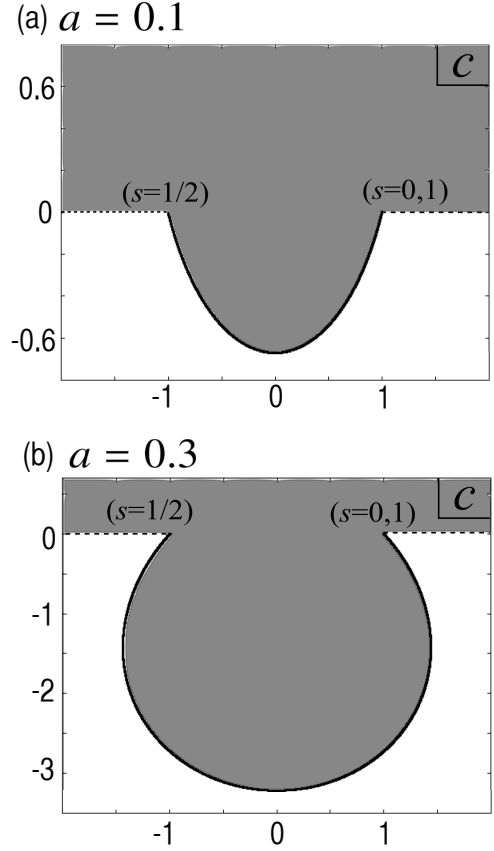}
            \caption{ Extension of the domain of $D(c)$.  The gray
              regions represent the domains of $D(c)$ for (a) $a=0.1$
              and (b) $a=0.3$.  The curve in the $c$ plane represents
              $U(y_{h}(s))$, where $y_{h}(s)$ is the deformed
              integration path for the homogeneous Rayleigh
              equation \eqref{eq:Reqh}.  With increasing $a$,
              the domain of $D(c)$ is extended to a larger area
              in the lower half $c$ plane, but the
              computational cost also increases.}
    \label{fig:sp-deform}
    \end{center}
\end{figure}

We choose the particular deformation function $h(s)$ in Eq. \eqref{eq:ysh} as
\begin{equation}
    \label{eq:hs}
    h(s) = a \sin 2\pi s, \quad a \geq 0.
\end{equation}
The extended domains of $D(c)$ are shown in Fig. \ref{fig:sp-deform}
for two different values of $a$.
We use $a$ large enough so that $\sigma_h$ is below $c$;
the computational cost increases exponentially
with increasing $a$.

By introducing the following functions of $s$:
\begin{equation}
    \label{eq:phiq}
    \varphi(s) = \phi(y_{h}(s)), \quad
    p(s) = q(y_{h}(s)),
\end{equation}
the homogeneous Rayleigh equation \eqref{eq:Reqh} is rewritten as
\begin{equation}
    \label{eq:Reqhs}
    \begin{split}
        \varphi_{\rm R}'' &- \frac{h'h''}{1 + h'^2} \varphi_{\rm R}' +
\frac{h''}{1 + h'^2} \varphi_{\rm I}' \\
        &- (2\pi \Gamma)^2 \left((1-h'^2) p_{\rm R} -2h' p_{\rm I} \right)
\varphi_{\rm R} \\
        &+ (2\pi \Gamma)^2 \left(2h' p_{\rm R} +(1-h'^2) p_{\rm I} \right)
\varphi_{\rm I }= 0, \\
        \varphi_{\rm I}'' &- \frac{h''}{1 + h'^2} \varphi_{\rm R}' -
\frac{h'h''}{1 + h'^2} \varphi_{\rm I}' \\
        &- (2\pi \Gamma)^2 \left(2h' p_{\rm R} +(1-h'^2) p_{\rm I}
\right)\varphi_{\rm R} \\
        &-(2\pi \Gamma)^2 \left((1-h'^2) p_{\rm R} -2h' p_{\rm I} \right)
\varphi_{\rm I }= 0,
    \end{split}
\end{equation}
where the subscripts ``R'' and ``I'' represent the real and imaginary parts
 respectively, and the prime denotes the derivative with
respect to $s$.  The continued solutions $\varphi_{1}$ and
$\varphi_{2}$ are computed by solving Eq. \eqref{eq:Reqhs} with the
boundary conditions
\begin{equation}
    \label{eq:bcs}
    \begin{cases}
        \varphi_{1, {\rm R}}(0) &= 1, \\
        \varphi_{1, {\rm I}}(0) &= 0,\\
        \varphi_{1, {\rm R}}'(0) &= 0,\\
        \varphi_{1, {\rm I}}'(0) &= 0,
    \end{cases}
    ~\textrm{and}\quad
    \begin{cases}
        \varphi_{2, {\rm R}}(0) &= 0,\\
        \varphi_{2, {\rm I}}(0) &= 0, \\
        \varphi_{2, {\rm R}}'(0) &= 2\pi \Gamma,\\
        \varphi_{2, {\rm I}}'(0) &= 2\pi \Gamma h'(0),
    \end{cases}
\end{equation}
respectively.
Solving Eq. \eqref{eq:Reqhs}, we obtain $\varphi_{1}(1)$ and 
$\varphi_{2}'(1)$, and the dispersion function is expressed as
\begin{equation}
    D(c) = 2 - \varphi_{1}(1) - \frac{\varphi_2'(1)}{2\pi\Gamma(1 +
ih'(1))},
\end{equation}
whose real and imaginary parts are respectively
\begin{equation}
    \label{eq:disp-rel}
    \begin{split}
        D_{\rm R}(c) &=
        2
        - \varphi_{1, \rm R}(1) - \frac{\varphi_{2, \rm R}(1)
        + h'(1) \varphi_{2, \rm I}(1)}{2\pi \Gamma(1 + h'(1)^2)},
        \\
        D_{\rm I}(c) &=
        - \varphi_{1, \rm I}(1) + \frac{h'(1) \varphi_{2, \rm R}(1)
        - \varphi_{2, \rm I}(1)}{2\pi \Gamma(1 + h'(1)^2)}.
    \end{split}
\end{equation}

\subsection{The main Landau pole and its bifurcation}

We now specialize the above computations to the initial perturbation
\begin{equation}
\label{eq:initial}
\omega_1(x,y,0) = -\epsilon \cos x,
\end{equation}
as in the numerics.
Using the analytical continuation of $D(c)$ into the lower half plane,
we numerically compute Landau poles, i.e. the roots of the equation
$D(c)=0$ with ${\rm Im}~c<0$. 
We are interested in
the traveling vortices, whose velocity is $c_{\rm R}$,
in resonance to the base flow, whose velocity $U(y)$ is in the range $[-1,1]$.
It is therefore natural to choose the analytical continuation of $D(c)$
defined by deforming the integral path as in Eq. \eqref{eq:hs}. 

\begin{figure}[t]
    \begin{center}
    \includegraphics[width=8cm]{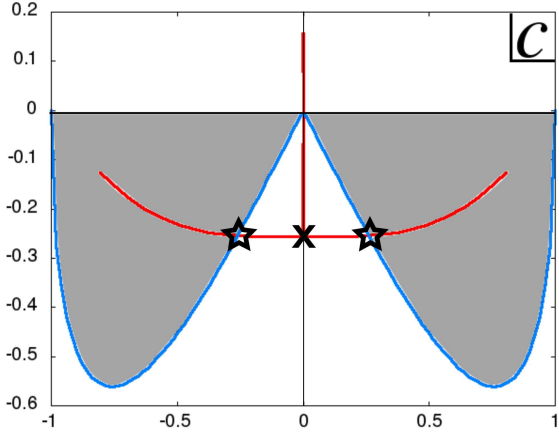}
    \caption{(color online)
The dominant root of the dispersion
          function (red line) and a favorable region for the existence of traveling vortices (gray region) on the complex $c$ plane.
When $\Gamma<1$, one dominant root is obtained 
on the upper half plane, 
where the base flow is unstable,
and the root is at the origin for $\Gamma=1$.
With increasing $\Gamma$,
the main root goes down to the lower half plane
along the imaginary $c$ axis, accompanied by another root going up the axis.
When $\Gamma=\Gamma_{\rm c} \simeq 1.06$,
these two roots collide at $c = -0.28 i$
and they bifurcate to the right and left directions.
The gray region is given by the inequality,
$c_{\rm I}^{2}<(1-c_{\rm R}^{2})(\arcsin(c_{\rm R}))^{2}$
coming from Eq. \eqref{eq:criterion}, and $c_{\rm I}<0$
required by the stability of the flow.
For later convenience, we marked the bifurcation point
corresponding $\Gamma=\Gamma_{\rm c}$
by the cross,
            and the point at which the main pole enters the gray region
            by thestars.
}
    \label{fig:bifurcation}
    \end{center}
\end{figure}

We look for the dominant Landau pole, i.e.,
the root of $D(c)$ with the
largest imaginary part; in a stable situation, it corresponds to the
slowest damping.  The variation of this pole as a function of the
aspect ratio $\Gamma$ is shown on Fig.~\ref{fig:bifurcation}. For
$\Gamma<1$, the flow is unstable.
With increasing $\Gamma$ above $1$,
the flow becomes stable, with the main Landau pole on the imaginary axis.
With further increasing $\Gamma$, the dominant Landau pole undergoes
a bifurcation at  an aspect ratio $\Gamma_{\rm c} \simeq 1.06$
and acquires a non-zero real part.
We will see that this bifurcation is crucial to understand the
appearance of a stationary dipole or of the four traveling vortices.

\section{The phase diagram}
\label{sec:phased}

Using the theoretical analyses above,
we construct the phase diagram on the $(\epsilon, \Gamma)$ plane,
which consists of zonal, dipole and oscillation phases \cite{HM11}.
The last oscillation phase corresponds to the
four traveling vortices.

\begin{figure}[t]
    \begin{center}
            \includegraphics[width=8cm]{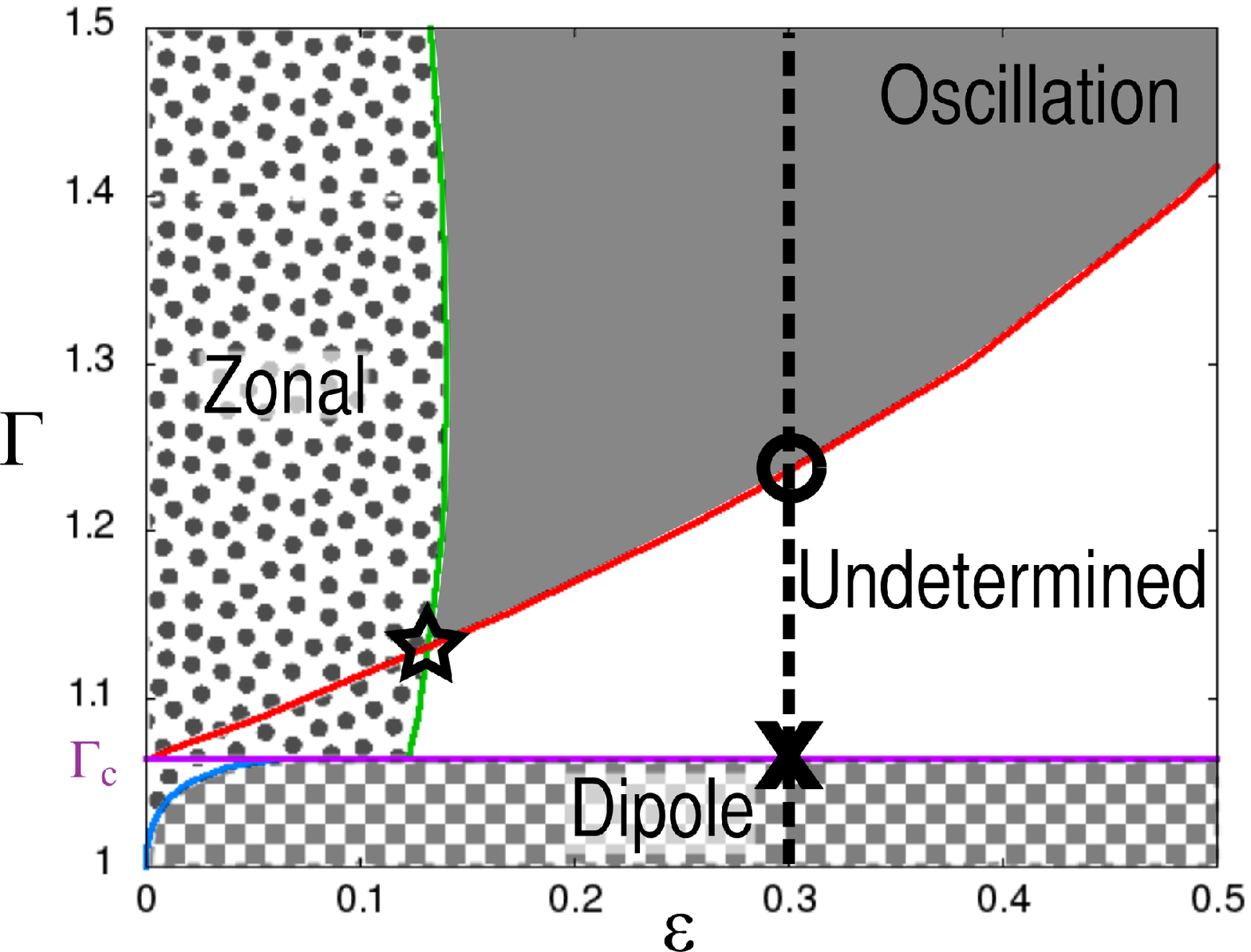}
            \includegraphics[width=8cm]{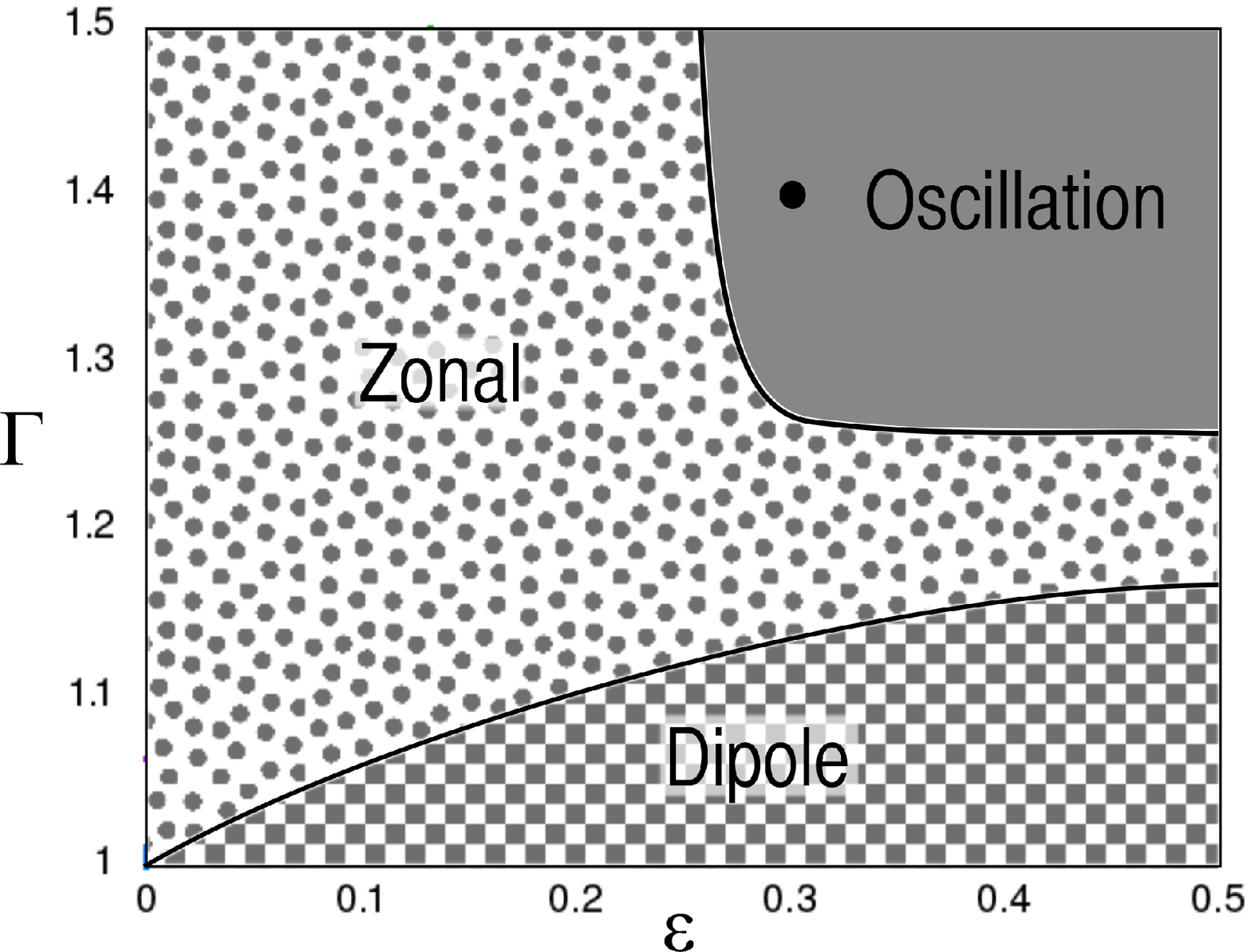}
    \caption{(color online)
          Phase diagram on the $(\epsilon,\Gamma)$ plane.
          The gray, polka-dotted and checkered regions
          correspond to the oscillatory, zonal and dipole phases,
          respectively.
          The upper panel is drawn by the present theory
          and the lower is a schematic picture of the phase diagram
          obtained with the numerical simulations by Morita \cite{HM11}.
          In the upper panel,
          the gray region is obtained from the inequality
          \eqref{eq:criterion}.
          The left green boundary of the gray region
          is given by $\epsilon = \Gamma c_{\rm I}^2/ \sqrt{1-c_{\rm R}^2}$
	(see Eq. \eqref{eq:cond1}),
          and the lower red by
          $\epsilon = \Gamma \sqrt{1-c_{\rm R}^{2}}(\arcsin c_{\rm R})^2$
	(see Eq. \eqref{eq:cond2}).
          The left blue boundary of the checkered region
          is given by $\epsilon=\Gamma c_{\rm I}^{2}$,
          and the upper purple by $\Gamma=\Gamma_{\rm c}$
	(see Eq. \eqref{eq:dipole}),
          whose value is determined at the bifurcation point
            marked by the cross in Fig. \ref{fig:bifurcation}.
            The star point corresponds to the star point
            of Fig. \ref{fig:bifurcation}.
          The broken line represents $\epsilon = 0.3$ on which
          the theoretical and numerical frequencies are compared
          in Fig. \ref{fig:comparison}.
          The points marked by the cross and the circle
            correspond to those in Fig. \ref{fig:comparison}, respectively.}
    \label{fig:phasediagram}
    \end{center}
\end{figure}

To obtain the region
on the $(\epsilon, \Gamma)$ plane
where the appearance of four traveling vortices is expected,
we combine criteria 1 and 2, that is Eqs. \eqref{eq:cond1} and \eqref{eq:cond2}
respectively, with the result on the dominant Landau pole.
Satisfying both criteria then reads:
\begin{equation}
    \label{eq:criterion}
    \dfrac{\Gamma c_{\rm I}^{2}}{\sqrt{1-c_{\rm R}^{2}}}
    < \epsilon
    < \Gamma \sqrt{1-c_{\rm R}^{2}} (\arcsin(c_{\rm R}))^{2}.
\end{equation}
The right inequality requires a Landau pole with non-zero $c_{\rm R}$,
which implies $\Gamma>\Gamma_{\rm c}$.
The region which satisfies Eq. \eqref{eq:criterion} on the 
$(\epsilon,\Gamma)$ plane is reported in Fig. \ref{fig:phasediagram},
which is qualitatively in good agreement with the numerically obtained
oscillatory region \cite{HM11}.  We remark that $|c_{\rm R}|$
increases when $\Gamma$ is increased above $\Gamma_{\rm c}$,
staying smaller than $1/\sqrt{2}$ in
the reported parameter region.  
This fact validates that we have omitted
the lower condition in Eq. \eqref{eq:2nd-necessary-condition-simple}
to derive the second necessary condition.

We expect to find a stationary dipole when criterion 1 is satisfied
and $c_{\rm R}=0$:
\begin{equation}
\label{eq:dipole}
    \Gamma c_{\rm I}^{2} < \epsilon \text{~~and~~} \Gamma<\Gamma_{\rm c}.
\end{equation}
The region satisfying these
requirements is highlighted in Fig. \ref{fig:phasediagram}.  This
theoretically estimated region is again in qualitative agreement
with the region where a stationary dipole emerges in numerical
simulations \cite{HM11}.

We expect a zonal flow when criterion 1 is not satisfied, that is
\begin{equation}
    \begin{split}
    \Gamma c_{\rm I}^2 &> \epsilon,
        \quad\textrm{for}\quad \Gamma < \Gamma_{\rm c}, \\
        %c_{\rm R} = 0,\\
    \dfrac{\Gamma c_{\rm I}^{2}}{\sqrt{1-c_{\rm R}^{2}}} &> \epsilon,
    \quad\textrm{for}\quad \Gamma > \Gamma_{\rm c}. 
        %c_{\rm R} \neq 0.
    \end{split}
\end{equation}
Physically, it means that nonlinearity is not dominant
and hence simple damping of the perturbation is expected.
In Fig. \ref{fig:phasediagram} the zonal region appears for small
$\epsilon$.

On Fig. \ref{fig:phasediagram}, there is a region where the theory
makes no prediction; indeed, the vortices that would be created by the
phenomenological mechanism considered above would be so close
one to another that they would strongly interact; the final fate of
the system is then out of scope of the present theory. We note that in
this region, zonal flows, dipoles and traveling vortices are 
observed numerically \cite{HM11}.

\section{Numerical tests}
\label{sec:num}
According to the phenomenological prediction
of the preceding sections,
the four traveling vortices run in the $x$ direction along the
four lines $y=y_{\ast}$ in
$\mathbb{T}^{2}=[0, 2\pi)\times [0,2\pi\Gamma)$,
where $y_{\ast}=\Gamma\arccos(\pm c_{\rm R})$, and their velocities are
$U(y_{\ast})=\pm c_{\rm R}$.  The period in the $x$ direction is $2\pi$;
hence the phenomenological theory predicts the frequency
$f_{\rm pt}=|c_{\rm R}|/2\pi$ for the vorticity field $\omega$.

We examine our phenomenological prediction by
numerically observing the $y$-positions of the four vortices 
and the oscillation frequency.
The initial condition of the vorticity field is:
\begin{equation}
    \label{eq:omega1-initial}
    \omega(x,y,0) = \omega_{0}(x,y) -\epsilon\cos x
\end{equation}
fixing the small parameter $\epsilon=0.3$.
In numerical simulations,
we add a hyper-viscous term $(-1)^{h+1}\nu\nabla^{2h}\omega$
to the right hand side of the 2D euler equation,
with $h=4$ and $\nu=2\cdot 10^{-18}$,
for stabilizing the numerical scheme,
and use the pseudo-spectral method with the resolution $256\times 256$.

\begin{figure}[t]
    \begin{center}
        \includegraphics[width=8cm]{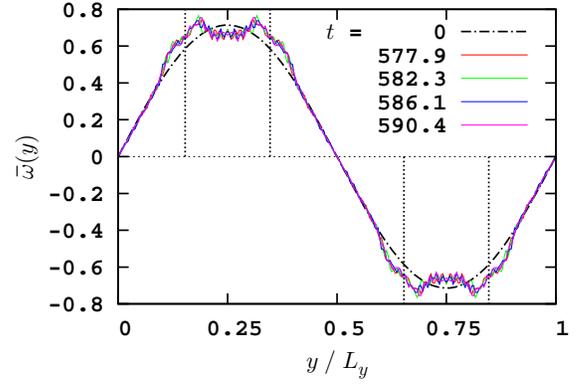}
    \caption{(color online)
      The averaged vorticity field $\bar{\omega}(y,t)$.
          Four bumps are observed, corresponding to four vortices
          in the long time regime.
          The vertical four dotted lines indicate the theoretical predictions
          for the $y$-positions of the vortices:
	$y/L_y=0.25\pm 0.097, 0.75\pm 0.097$.
          $\Gamma=1.4$ and $\epsilon=0.3$.}
    \label{fig:baromega}
    \end{center}
\end{figure}

First, we observe
the $y$-positions by computing
the averaged vorticity field $\bar{\omega}(y)$ defined by
\begin{equation}
    \bar{\omega}(y,t) = \dfrac{1}{2\pi} \int_{0}^{2\pi} \omega(x,y,t) dx.
\end{equation}
At the initial time $t=0$,
the averaged vorticity field is $\omega_{0}(y)=\sin(y/\Gamma)/\Gamma$,
and it evolves in time.
Asymptotically, we observe four bumps in $\bar{\omega}(y)$
as shown in Fig. \ref{fig:baromega}
for $\Gamma=1.4$ and $\epsilon=0.3$.
The theoretical prediction is in good agreement with the numerically 
computed bump positions.

\begin{figure}[t]
    \begin{center}
    \includegraphics[width=8cm]{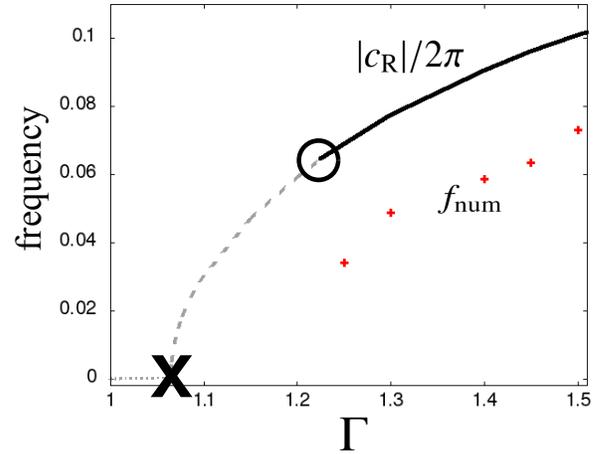}
    \caption{(color online) $\Gamma$ dependence of the frequencies
          obtained from the theoretical prediction (bold curve)
          and the numerical simulations (red points).
          $\epsilon=0.3$.
          Clear oscillation is not found by numerical simulations 
          in the small $\Gamma$ region, 
          $\Gamma \alt 1.2$. 
	The points marked by the cross and the circle correspond 
	to those in Fig. \ref{fig:phasediagram}, respectively.}
    \label{fig:comparison}
    \end{center}
\end{figure}

Next, we obtain the frequency
by computing the power spectra of
the $(1,0)$ Fourier mode of $\omega(x,y,t)$ defined by
\begin{equation}
    \label{eq:observable}
    Z(t) = - {\rm Re}~ \dfrac{1}{(2\pi)^{2}\Gamma} \iint_{\mathbb{T}^2}
\omega(x,y,t) e^{-ix} dxdy.
\end{equation}
As shown in Fig. \ref{fig:comparison}, the dependence on $\Gamma$
of $f_{\rm num}$ is in qualitative agreement with the prediction, in the
sense
that the frequency increases as $\Gamma$ increases.  However, it is
not in good agreement quantitatively.
One possible explanation for the quantitative discrepancy in frequency
is that $\epsilon=0.3$ is too large
to be considered a small perturbation to the base flow.
The present theory is based on the linear analysis of the Euler equation.
It gives good predictions qualitatively, but quantitatively,
nonlinear effects may kick in for rather large $\epsilon$.

\section{Summary and discussions}
\label{sec:sum}
Inspired by previous works on the Vlasov equation
\cite{TMO65,JPH91,MB95,CL98,JB09}, we have considered i) the balance
between Landau damping and nonlinear effects
and ii) the non-overlapping criterion between nonlinear structures, 
in order to analyze
the formation of a stationary dipole and traveling vortices in
the 2D Euler equation starting from the perturbed stable Kolmogorov flow.
The detailed linear analysis provides
information on this nonlinear dynamical phenomenon.
We emphasize that the bifurcation of the dominant Landau pole
as the aspect ratio is varied plays a crucial
role in the theory.
Since this kind of
bifurcation is probably not a special feature of the Kolmogorov flow,
we expect that a similar analysis, predicting the appearance, or not,
of a dipole or traveling vortices, would be fruitful for other base flows.

We note that our prediction is in qualitative, though not quantitative, 
agreement with the numerical results of
Ref. \cite{HM11}, not only for the shape of the phase diagram, but also 
for the vortex frequency.
The quantitative discrepancy for the phase diagram is rather natural
since we have performed phenomenological order-of-magnitude estimates
and not taken the prefactors into consideration.
The reason for the quantitative disagreement in
the frequency is probably that the size of the perturbation
$\epsilon$ used in the numerical simulation is large enough to
trigger nonlinear effects that we do not take into account,
while such a large $\epsilon$ is necessary to realize the oscillating 
phase numerically.

\acknowledgments
The authors are grateful to F. Bouchet for fruitful discussions
and comments on this study.
SO thanks for the hospitality of Universit\'e de Nice Sophia-Antipolis
and acknowledges the support of
the JSPS Research Fellowships for Young Scientists
(Grant No. 254728).
This work was supported by the ANR-09-JCJC-009401 INTERLOP project.
YYY acknowledges the support of a Grant-in-Aid for Scientific Research (C)
23560069.


\begin{thebibliography}{99}

      \bibitem{FB12}
    F. Bouchet and A. Venaille,
    Phys. Rep. {\bf 515}, 227 (2012).

      \bibitem{NJB13}
    N. J. Balmforth, P. J. Morrison, and J. L. Thiffeault,
    arXiv:1303.0065.

      \bibitem{PSM88}
    P. S. Marcus,
    Nature {\bf 331}, 693 (1988).

      \bibitem{BM11}
    J. B. Marston,
    Physics {\bf 4}, 20 (2011).

    % 5
      \bibitem{AV11}
    A. Venaille and F. Bouchet,
    J. Phys. Ocean. {\bf 41}, 1860 (2011).

      \bibitem{WT80}
    W. Thomson,
    Nature {\bf 23}, 45 (1880).

      \bibitem{HL32}
    H. Lamb,
    Hydrodynamics, 6th edition (Cambridge University Press, Cambridge,
1932).

      \bibitem{JTS67}
    J. T. Stuart,
    J. Fluid Mech. {\bf 29}, 417 (1967).

    % 10
      \bibitem{HMR86}
    D. Holm, J. Marsden, and T. Ratiu. in: Nonlinear
    systems of partial differential equations in applied
    mathematics,  Part 2, 171-186.
    Series: Lectures in Appl. Math., vol. 23 (1986).

      \bibitem{ZL11}
    Z. Lin and C. Zeng,
    Arch. Rational Mech. Anal. {\bf 200}, 1075 (2011).

      \bibitem{LO49}
    L. Onsager,
    Nuovo Cimento. Supple. {\bf 6}, 279 (1949).

      \bibitem{GLE06}
    G. L. Eyink and K. R. Sreenivasan,
    Rev. Mod. Phys. {\bf 78}, 87 (2006).

      \bibitem{JM90}
    J. Miller,
    Phys. Rev. Lett. {\bf 65}, 2137 (1990) .

    % 15
      \bibitem{JM92}
    J. Miller, P. B. Weichman, and M. C. Cross,
    Phys. Rev. A {\bf 45}, 2328 (1992).

      \bibitem{RR91}
    R. Robert,
    J. Stat. Phys. {\bf 65}, 531 (1991).

      \bibitem{RRJS91}
    R. Robert and J. Sommeria,
    J. Fluids. Mech. {\bf 229}, 291 (1991).

      \bibitem{HM11}
    H. Morita, arXiv:1103.1140.

      \bibitem{LDL45}
    L. D. Landau,
    J. Phys. U.S.S.R. {\bf 10}, 25 (1946);
    Collected papers of L. D. Landau
    edited by D. T. Haar
    (Pergamon Press, Oxford,1965).

    % 20
      \bibitem{TMO65}
    T. M. O'Neil,
    Phys. Fluids {\bf 8}, 2255 (1965).

      \bibitem{GM97} G. Manfredi,
    Phys. Rev. Lett. {\bf 79}, 2815 (1997).

      \bibitem{MB00} M. Brunetti, F. Califano, and F. Pegoraro,
    Phys. Rev. E {\bf 62}, 4109 (2000).

      \bibitem{CL98}
    C. Lancellotti and J. J. Dorning,
    Phys. Rev. Lett. {\bf 81}, 5137 (1998);
    Phys. Rev. E {\bf 68},  026406 (2003);
    Trans. Th. Stat. Phys. {\bf 38}, 1 (2009).

      \bibitem{JB09}
    J. Barr\'e and Y. Y. Yamaguchi,
    Phys. Rev. E {\bf 79}, 036208 (2009).

    % 25
      \bibitem{MB95}
    M. Buchanan and J. J. Dorning,
    Phys. Rev. Lett. {\bf 70}, 3732 (1993);
    Phys. Rev. E {\bf 52} 3015 (1996).

      \bibitem{RLS97}
    R. L. Spencer and S. N. Rasband,
    Phys. Plasmas {\bf 4}, 53 (1997).

      \bibitem{FB10}
    F. Bouchet and H. Morita,
    Physica D {\bf 239}, 948 (2010).

      \bibitem{NRC95}
    N. R. Corngold,
    Phys. Plasmas {\bf 2}, 620 (1995).

      \bibitem{RJB70}
    R. J. Briggs, J. D. Daugherty, and R. H. Levy,
    Phys. Fluids {\bf 13}, 421 (1970).

    % 30
      \bibitem{JDC89}
    J. D. Crawford and P. D. Hislop,
    Ann. Phys. {\bf 189}, 265 (1989).

      \bibitem{PDH89}
    P. D. Hislop and J. D. Crawford,
    J. Math. Phys. {\bf 30}, 2819 (1989).

      \bibitem{JPH91}
    J. P. Holloway and J. J. Dorning,
    Phys. Rev. A {\bf 44}, 3856 (1991).



\end{thebibliography}
\end{document}